\documentclass[10pt,letterpaper]{article}
\usepackage[margin=1.35in]{geometry}
\usepackage{setspace}

\usepackage{epsfig, amsthm, amsmath, amssymb, latexsym, xpatch, xcolor, mathtools, verbatim, cancel}
\usepackage[titletoc]{appendix}
\usepackage[normalem]{ulem} % strike out \sout{...} % for better underline, use \uline{...}
\usepackage{tabularx}
\usepackage{stmaryrd} % for linguistics symbols
\usepackage[margin=1cm]{caption}
\usepackage{float} % to make figure exactly here
\usepackage{wrapfig}
\usepackage{multicol} % for two-column layout
\usepackage{hyperref}

\newcommand{\VV}{\mathbf{V}}
\newcommand{\UU}{\mathbf{U}}
\newcommand{\vv}{\mathbf{v}}

\newcommand{\op}{\begin{itemize}}
\newcommand{\ed}{\end{itemize}}

\newcommand{\opp}{\begin{quote}}
\newcommand{\edd}{\end{quote}}

\newcommand{\ope}{\begin{enumerate}}
\newcommand{\ede}{\end{enumerate}}

\newcommand{\xm}{\item[]}
\newcommand{\im}{\item}

\newcommand{\PP}{\mathbb{P}}
\newcommand{\given}{\mbox{$\,|\,$}}
\newcommand{\Given}{\mbox{$\;\big|\;$}}
\newcommand{\GGiven}{\mbox{$\;\big| \!\big|\;$}}

\newcommand{\Cure}{\mathit{Cure}}
\newcommand{\Take}{\mathit{Take}}
\newcommand{\take}{\mathit{take}}

\newcommand{\Individual}{\mathit{Indiv}}

\newcommand{\Assign}{\mathit{Assign}}
\newcommand{\assign}{\mathit{assign}}

\definecolor{cardinal}{rgb}{0.8, 0.0, 0.0}
\definecolor{deepblue}{rgb}{0.0, 0.2, 0.8}

\makeatletter
\def\th@plain{%
  \thm@notefont{}% same as heading font
  \itshape % body font
}
\makeatother

\theoremstyle{plain}
\newtheorem{theorem}{Theorem}

\newtheorem*{lemma*}{Lemma}

\title{Never Too LATE: A Fully Stochastic Update to the Potential Outcome Framework}
\author{Hanti Lin \\[0.5em] University of California, Davis \\{ika@ucdavis.edu}}

\date{}

\begin{document}

\maketitle

\begin{abstract} \noindent In the classic potential outcome framework, the local average treatment effect (LATE) and its identification via an instrumental variable are stated in a deterministic setting at the individual level: each individual has settled potential outcomes such as ``cured if treated''. Several authors have proposed working instead with \emph{stochastic} potential outcomes---counterfactual probabilities of the form ``the chance of being cured if treated''---but the integration of stochastic potential outcomes with the LATE machinery raises an issue. It is a metaphysical issue: in a stochastic setting, the standard joint-probability definitions of compliers and the LATE assume what I will call the \emph{unique-parallel-universe view}, which asserts that, in any genuinely possible state of the world, every counterfactual condition settles a unique determinate outcome even when the underlying causal disposition is irreducibly chancy. The statistician Dawid (2000) doubts the plausibility of this view; the philosopher Lewis (1973) develops a reductio argument against it. I propose a fully stochastic update to the Rubin causal model that drops the assumption of the unique-parallel-universe view: stochastic potential outcomes are introduced as Bernoulli parameters in their own (small) probability spaces, and are connected to observables via the factorization rule of a causal Bayes net. Within this framework, I define a Degree-of-compliance-weighted Average Treatment Effect (DATE) and prove that, under assumptions analogous to those used for the LATE but rewritten for the fully stochastic setting, the DATE equals the usual IV estimand. The classic LATE identification result emerges as a deterministic special case. Existing IV practice can therefore be reinterpreted: it has been estimating the DATE all along, in a general stochastic setting, without assuming the unique-parallel-universe view.
\end{abstract}

%The Rubin causal model, together with its application to the instrumental variable (IV) estimation of the local average treatment effect (LATE), is known to make a strong, determinist assumption (Dawid 2000). In fact, a parallel concern was raised by Lewis (1973) in philosophy, directed at the determinist assumption underlying Stalnaker's (1968) semantics of counterfactuals. This paper responses to those challenges by developing a fully stochastic update to the Rubin causal model, leading to a generalized result for identification of the LATE. This is achieved by incorporating an idea from Robins \& Greenland (1991) into a causal Bayes net (rather than Pearl's (2009) nonparametric structural equation models). The use of a causal Bayes net, together with its underlying DAG, makes it very easy to state assumptions and prove the new identification result in a fully stochastic setting.

%\newpage
%\addcontentsline{toc}{section}{Table of Contents}
\tableofcontents
%\newpage

\section{Introduction}\label{sec_intro}

Causal inference from observational data often uses an instrumental variable to identify the local average treatment effect (LATE) of a treatment on a subpopulation of compliers (Imbens and Angrist 1994; Angrist, Imbens, and Rubin 1996). The classic LATE theorem is stated within the potential outcome framework, in which each individual $i$ has settled, deterministic potential outcomes such as $\Cure_i^{\take=1}$ (whether $i$ would be cured if $i$ took the treatment) and $\Take_i^{\assign=1}$ (whether $i$ would take the treatment if assigned to it). The four standard compliance types---complier, never-taker, always-taker, defier---are defined by combinations of such outcomes, and, under suitable assumptions, the LATE is identified with a familiar ratio of expectations of observables.

But many real causal dispositions are irreducibly chancy. A vaccine raises the probability of immunity rather than guaranteeing it; an offered treatment raises the probability of compliance rather than determining it. To accommodate such cases, a stream of work in epidemiology---Greenland (1987), Robins and Greenland (1989, 2000), and VanderWeele and Robins (2012)---has proposed working with \emph{stochastic potential outcomes}: counterfactual probabilities of the form ``the probability $p$ such that, if $i$ took the treatment, $i$ would have a chance $p$ of being cured'', rather than counterfactual binary values of the form ``whether $i$ would be cured if $i$ took the treatment''. Stochastic potential outcomes are the natural object of interest when the underlying world is genuinely indeterministic. But their integration with the LATE machinery has raised a foundational issue.

The issue is metaphysical, with technical and mathematical ramifications. The classic work on the identification of the local average treatment effect for the subpopulation of compliers rests on a joint probability distribution over observable and counterfactual variables---expressions such as 
$$
\mathbb{P}\bigl(\Take^{\assign=0} = 0,\ \Take^{\assign=1} = 1\bigr)
$$
for the probability of sampling a complier. Once these expressions are interpreted stochastically, they tacitly assume what I will call the \emph{unique-parallel-universe view}: the substantive Laplacian thesis that, under any counterfactual condition, there is a unique determinate fact of the matter about the outcome---even when the underlying causal disposition is genuinely chancy. As we will see below, the statistician Dawid (2000) doubts the plausibility of this view. In a different context, the philosopher Lewis (1973) even develops a reductio argument against that view. The debate persists in today's philosophical community, among metaphysicians and philosophers of language (more on this below). Whichever way the debate goes, the standard joint-probability formulations are not metaphysically neutral.

My aim is to develop a stochastic generalization of the LATE identification result that {\em drops} the assumption of the unique-parallel-universe view and does {\em not} take a stance on this metaphysical debate. The proposal is to accept only a much weaker commonsense metaphysics. But this metaphysical caution poses a technical challenge: unlike the classic setup, I cannot use the standard machinery---a single joint probability distribution over observables and counterfactual variables---to state definitions, theorems, or proofs. This is essentially {\em the} challenge posed by Dawid (2000).  

To take up the challenge, my strategy is to combine stochastic potential outcomes and the machinery of causal Bayes nets. The idea is to assume a bridge principle that connects stochastic potential outcomes to the conditional probabilities of observables (given their direct causes) that appear in the factorization rule of an appropriate causal Bayes net. Such a factorization rule basically expresses what is known as the causal Markov assumption. 

Within this new framework, we can define a new quantity for estimation: the Degree-of-compliance-weighted Average Treatment Effect (DATE). It can then be proved that, under assumptions analogous to those used for the LATE, the DATE equals the usual IV estimand. In the deterministic special case, the DATE reduces to the LATE, so the classic identification result is a corollary. The upshot is that existing IV practice can be reinterpreted: previously understood as estimating the LATE in a deterministic setting, it has actually been estimating the DATE all along---in a fully stochastic setting, with the deterministic setting as a special, degenerate case, and without taking on the unique-parallel-universe view.

The paper is organized as follows. Section~\ref{sec:metaphysical-issue} states and sharpens the metaphysical issue, contrasting the metaphysical assumption of the unique-parallel-universe view with a much weaker commonsense alternative and rehearsing Lewis's (1973) reductio. Section~\ref{sec_late} (``Going Stochastic'') implements the strategy sketched above in three steps and reports the main identification theorem I proved elsewhere (in a paper written for philosophy, focusing on logical rather than metaphysical issues), but for completeness a proof is included in the Appendix. In Section~\ref{sec_challenges}, I defend the choice to use a graphical model rather than the graph-less Rubin causal model alone, addressing the concern expressed by Imbens (2020). Among the graphical causal models, my choice is causal Bayes nets rather than endogenously deterministic structural equation models, and this is needed to do without assuming the unique-parallel-universe view. This choice is also defended, addressing Pearl's worries about the limitations of causal Bayes nets. Section~\ref{sec_conclusion} concludes by abstracting a general three-step recipe for stochasticizing the classic potential outcome framework.

\section{Some Metaphysics}
\label{sec:metaphysical-issue}

Before we move to a stochastic generalization of the potential outcome framework, it pays to pause on
\opp 
{\em A Foundational Question}: What are the possibilities---genuine possibilities---to which we want to assign probabilities? 
\edd 
Indeed, we have to use probability measures, and each presupposes a sample space---a space of possibilities. But this requires us to fix what counts as a genuine possibility---and, in particular, how \emph{finely} possibilities are individuated. 

\subsection{Coarse-Grained vs. Fine-Grained}

To warm up, consider an ordinary, non-causal example. Consider the following coarse-grained possibility:
\opp
\emph{It rains.}
\edd
Starting from this, we can refine it by adding more descriptions, thereby constructing more fine-grained possibilities, such as
\opp
\emph{It rains, and the bus comes on time.}
\edd
and
\opp
\emph{It rains, and the bus comes late.}
\edd
The idea is that a coarse-grained possibility may be divided up by adding one declarative sentence or another. This looks innocuous and familiar. But, in the stochastic setting, analogous moves can carry hidden metaphysical baggage. Let me explain.

\begin{figure}[ht]
	\centering \includegraphics[width=0.4\textwidth]{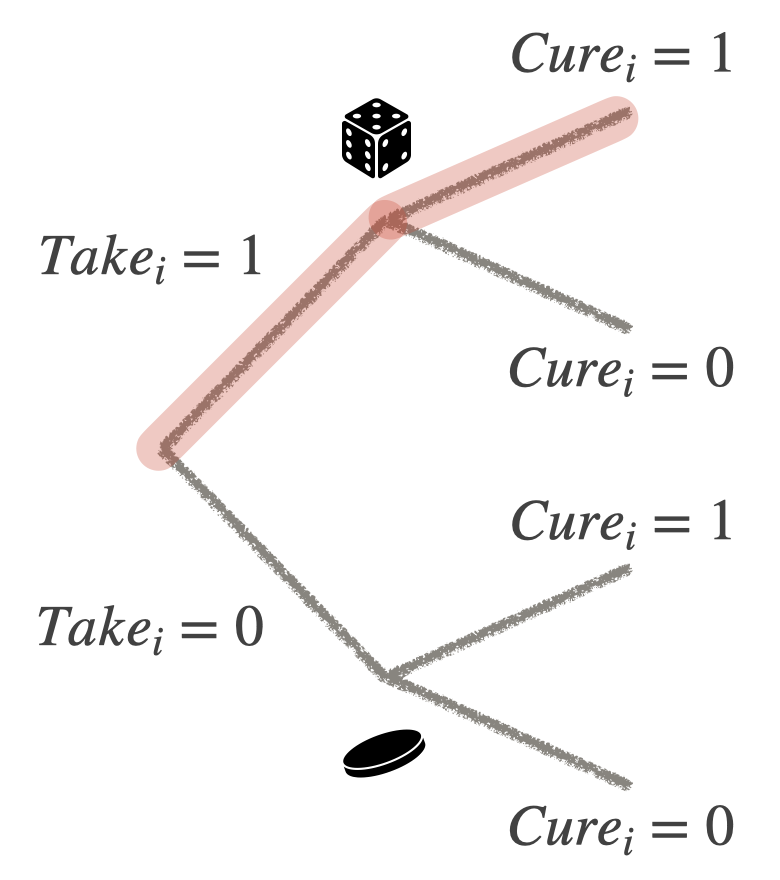}
	\caption{\em A coarse-grained possible state of the world, where the red branch represents what actually happens in that state}
	\label{fig-coarse}
	\end{figure}

Consider a study of a treatment whose effect is estimated in part in terms of whether each sampled individual $i$ is cured. Suppose we are working in a stochastic setting, so we allow for irreducibly chancy counterfactuals. Consider then the following possibility, which is also pictorially presented in figure \ref{fig-coarse}:
\opp
{\bf A Coarse-Grained State of the World}
\op
\im The sampled individual $i$ takes the treatment and is cured. (\emph{This is what actually happens in this possible state of the world, depicted as the red branch in figure \ref{fig-coarse}.})
\im If $i$ had not taken the treatment, $i$ would have had a $50\%$ chance of being cured. (\emph{This is a stochastic counterfactual, and the $50\%$ chance therein is a stochastic potential outcome under the stated counterfactual condition, represented by a coin at the lower branching point in figure \ref{fig-coarse}.})
\ed
\edd
The second sentence says that something stochastic holds in this state of the world: under the counterfactual antecedent $\take = 0$, $i$'s being cured is governed by a chance of $0.5$ as if tossing a fair coin, and nothing further is said about which way it would actually have gone---after all, ``that fair coin'' is not actually tossed.

Now let us try to refine this possibility by adding one more sentence.

\begin{figure}[ht]
	\centering \includegraphics[width=0.4\textwidth]{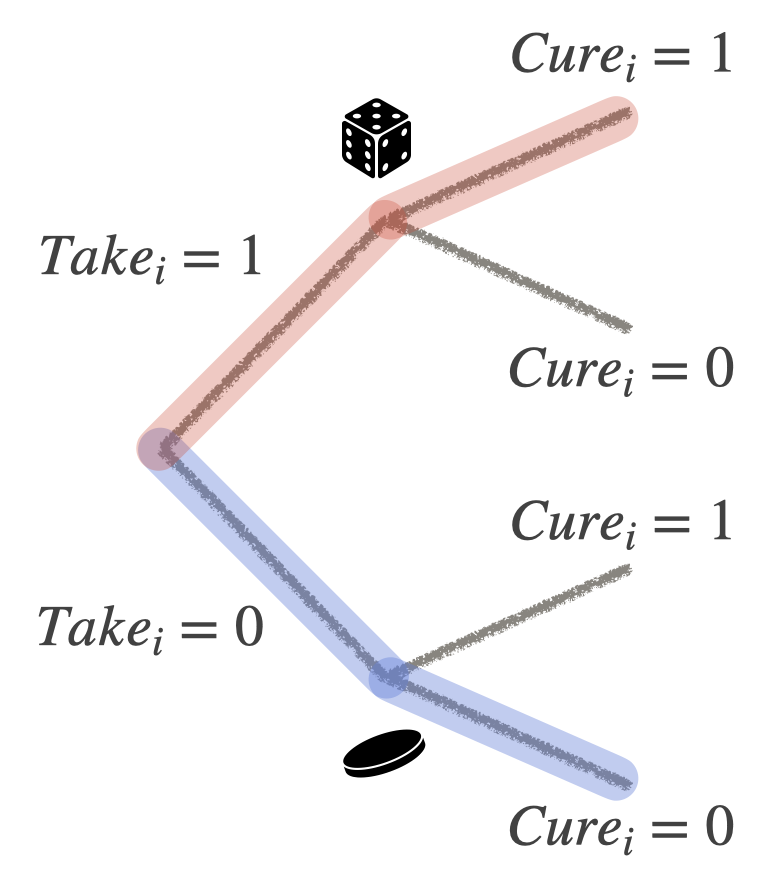}
	\caption{\em A fine-grained possibility, where the red branch represents the actual universe, and the blue branch represents the unique parallel universe under the counterfactual condition of not taking the treatment}
	\label{fig-fine}
	\end{figure}

\opp 
{\bf A Fine-Grained State of the World}
\op
\im The sampled individual $i$ takes the treatment and is cured.
\im If $i$ had not taken the treatment, $i$ would have had a $50\%$ chance of being cured. (\emph{To reiterate, this is a stochastic counterfactual.})
\im If $i$ had not taken the treatment, $i$ would not be cured. (\emph{This is a simple, non-stochastic counterfactual.})
\ed
\edd
A similar fine-grained possibility can of course be constructed by replacing the last `would not' with `would'---but let us focus on this one. Such a fine-grained possibility can be pictorially presented as in figure \ref{fig-fine}. 
There, the old, red branch still expresses what actually happens in a state of the world, whereas the newly added, blue one expresses a potential outcome in the standard formalism of the Rubin causal model, written $\Cure^{\take = 0} = 0$. To take such a fine-grained state of the world seriously as a genuine possibility is to accept the following metaphysical view:
\opp 
{\bf The Unique-Parallel-Universe (Laplacian) View}.
Genuine possibilities include fine-grained possibilities that have the following three elements simultaneously:
\op
\im a \emph{factual} universe, specifying what actually happens, represented by the red branch in figure \ref{fig-fine}; 
\im for each possible intervention of interest, a \emph{unique counterfactual universe} in which the outcome under that intervention is fully \emph{settled}---expressed by a simple, non-stochastic counterfactual, and represented by the blue branch in figure \ref{fig-fine};
\im together with some stochastic counterfactuals (such as ``if not treated, $i$ would have a $50\%$ chance of being cured'').
\ed 
\edd
The view is ``Laplacian'' in spirit: for every counterfactual antecedent of interest, there is supposed to be a unique determinate fact of the matter about the outcome---a unique counterfactual universe running in parallel with the factual one---\emph{even when} the agent's causal disposition under that antecedent is genuinely chancy. If there are parallel universes, there should be multiple ones---why should there be a unique one? 

To see how strong the unique-parallel-universe view is, compare it with a weaker alternative, which I will call the commonsense view:
\opp 
{\bf The Commonsense View}

Genuine possibilities include \emph{at least} coarse-grained ones, including the possibilities specified just by:
\op
\im a factual universe,
\im together with some stochastic counterfactuals.
\ed
\edd
On this commonsense view, we are not required to postulate, for every counterfactual antecedent, a unique fully-settled outcome to sit alongside the stochastic counterfactual. Stochasticity all the way down is permitted: under the antecedent $\take = 0$, it is permitted that the world is simply such that there is a $0.5$ chance of cure, full stop, with no further fact about ``which way it would actually have gone''. The commonsense view is officially silent about whether \emph{additional}, more fine-grained possibilities are genuine possibilities; it just doesn't demand them. The unique-parallel-universe view, by contrast, insists on the further structure.

That said, there may be something nice about the unique-parallel-universe view. It may seem convenient if we could place such fine-grained possibilities in the sample space for our probabilistic theory of causal inference. Two benefits would follow. Benefit 1: the fine-grained possibilities use the familiar non-stochastic potential outcomes of the standard Rubin causal model, so the existing definitions of important concepts---such as compliers, never-takers, the average treatment effect (ATE), and the local average treatment effect (LATE) over the subpopulation of compliers---can be carried over essentially unchanged. Benefit 2: the probabilities seem to combine in a perfectly natural way. If a probability $p$ is assigned to the coarse-grained possibility (which says only that the chance of cure under $\take = 0$ is $0.5$), then the fine-grained refinement that further specifies $\Cure^{\take = 0} = 0$ can be assigned probability $p \cdot (1 - 0.5) = 0.5\, p$, since the stochastic counterfactual itself awards probability $1 - 0.5$ to ``not cured''. On the basis of such fine-grained possibilities one can take further steps to construct a single joint probability measure over factual and counterfactual variables alike, retain the familiar Rubin-style definitions, and recover the classical machinery---including the standard definition of the LATE. Then perhaps there is no need to prove a new theorem and it suffices to reinterpret the classic identification result in the stochastic setting. 

The above are the benefits available if we are really willing to put fine-grained possibilities in the sample space. But, as we will see shortly, there is a price to pay---it makes a substantive metaphysical assumption: the unique-parallel-universe view. The statistician Dawid (2000) already voices a concern of roughly this kind, though in a slightly different wording (such as `determinism' and `fatalism'), without explicitly referring to the issue of granularity. 

Worse, the philosopher Lewis (1973) even develops a {\em reductio} argument against the unique-parallel-universe view, though he does it in a quite different context---in logic and natural language semantics rather than in metaphysics or causal inference.\footnote{Lewis would refer to the unique-parallel-universe assumption as the {\em unique-closest-world} assumption in the context of possible world semantics.} Adapted to our setting, Lewis's argument runs as follows.

\subsection{Lewis's Reductio Argument}

Suppose, \emph{for reductio}, that the following three sentences are jointly true at some possible state of the world:
\op
\im[$(1)$] The sampled individual $i$ is assigned to control and does not take the treatment. 
\im[$(2)$] If $i$ were assigned to treatment, $i$ would have a $50\%$ chance of taking the treatment. 
\im[$(3)$] If $i$ were assigned to treatment, $i$ would take the treatment.
\ed
This is precisely a fine-grained possibility of the kind the unique-parallel-universe assumption is committed to: a factual sentence (1), a stochastic counterfactual (2), and a simple counterfactual (3) settling the outcome under an intervention.

Now focus on (2) and (3), as the contradiction to be derived below actually has nothing to do with (1). Combining (2) with Lewis's key semantic principle that ``would have a less-than-one chance'' entails ``might not'', we have:
\op
\im[$(4)$] If $i$ were assigned to treatment, $i$ \emph{might not} take the treatment.
\ed
Conjoining the two counterfactuals in (3) and (4), we then have:
\op
\im[$(5)$] If $i$ were assigned to treatment, $i$ would take the treatment \emph{and} might not take it.
\ed
But this is a contradiction! 

So the declarative sentences (1)--(3) are jointly inconsistent and, thus, fail to define a genuine possibility. If we really want to assign to it a probability, it can only be a trivial one: probability zero. 

The argument turns on a single semantic principle---``would have a less-than-one chance'' entails ``might not''. By Lewis's lights, then, the unique-parallel-universe view is not merely a strong assumption but an incoherent one. The contemporary debate over this argument in the philosophical community is by no means settled: there are defenders of the unique-parallel-universe view (e.g.,\ Stalnaker (1981); Williams (2010); Pearl (2009); Santorio (2025)) and critics (e.g.,\ Lewis (1973); H{\'a}jek (2014, manuscript)). What matters for present purposes is that the issue is live: the unique-parallel-universe assumption is at the very least substantive, and arguably untenable. 

I propose that, if possible, we should at least try to develop a theory of causal inference that takes no stance on this metaphysical issue---neither assuming the unique-parallel-universe view, nor assuming the falsity of that view.

\subsection{Compliers and Joint Distributions in the Stochastic Setting}

To see why this is far from an arcane matter in theoretical philosophy, consider how the standard categories of the classic LATE framework---compliers, never-takers, always-takers, defiers---behave once we move to a stochastic setting. Start with a quick review:
	\opp 
	{\bf Classic Definition of Compliers}. An individual $i$ is said to be a complier in the classic sense if $i$ meets the following two conditions: 
	\op
	\im $i$ would not take the treatment if assigned to control (written $\Take^{\assign=0}_i = 0$);
	\im $i$ would take the treatment if assigned to it (written $\Take^{\assign=1}_i = 1$).
	\ed 
	\edd 
In the stochastic setting, however, suppose that for some sampled $i$ we have:
\opp 
{\bf A Coarse-Grained State of the World}
\op
\im $i$ is assigned to control and does not take the treatment;
\im if $i$ were assigned to treatment, $i$ would have only a $50\%$ chance of taking it.
\ed
\edd 
Calling $i$ a ``complier'' in the standard sense now requires adding a further description on top of the stochastic counterfactual---namely, ``if $i$ were assigned to treatment, $i$ would take it''. Then we have:
\opp 
{\bf A Fine-Grained State of the World}
\\{\bf (A Classic Complier in the Stochastic Setting)}
\op
\im $i$ is assigned to control and does not take the treatment;
\im if $i$ were assigned to treatment, $i$ would have only a $50\%$ chance of taking it;
\im if $i$ were assigned to treatment, $i$ would take it.
\ed
\edd 
That is precisely a fine-grained possibility of the kind Lewis attacks---we can run Lewis's argument again here. 

The point becomes especially vivid once we look at the joint probabilities that standardly define the subpopulation of compliers. In the classic potential outcome framework, one writes
\opp
$\mathbb{P}\bigl(\Take^{\assign=0} = 0,\ \Take^{\assign=1} = 1\bigr)$
\edd
for the probability of sampling a complier, and
\opp
$\mathbb{P}\bigl(\Assign = 0,\ \Take = 0,\ \Take^{\assign=1} = 1\bigr)$
\edd
for the probability of sampling a complier who ends up assigned to control and not taking the treatment. These expressions look innocuous, and in the classic, deterministic setting they are. But when they are used in the \emph{stochastic} setting, they are joint probabilities over factual variables and counterfactual variables together, and they make sense only if the corresponding fine-grained possibilities exist---only, that is, if the unique-parallel-universe assumption holds. 

But do we really want to, or need to, assume this strong metaphysical view? No, I claim. The challenge for me, taken up in the next section, is to develop a stochastic theory of (a suitable generalization of) the LATE that does justice to the commonsense view while remaining neutral on, or even rejecting, the unique-parallel-universe view.

\section{Going Stochastic}\label{sec_late}

Now let me develop a stochastic update to the potential outcome framework, which involves three steps. Here is the starting point:
	\opp  
	{\bf Step 1 (Identify Potential Outcomes).} Identify the (individual-level) potential outcomes used in the Rubin causal model: $\Cure_i^{\take = 1}, \Cure_i^{\take = 0}, \Take_i^{\assign = 1}$, and $\Take_i^{\assign = 0}$, where the index $i$ ranges over the individuals in the population.
	\edd 	
Those potential outcomes will be replaced by their stochastic counterparts. Compare the following two.
	\op 
	\xm \emph{Potential Outcome}: $\Cure_i^{\take = 1}$, which denotes the medical outcome that $i$ would have if $i$ took the treatment.
	\xm \emph{Stochastic Potential Outcome}: $\kappa_i^{\take=1}$, which denotes the (unknown) probability $p$ such that, if $i$ took the treatment, $i$ would have a chance $p$ of being cured.  
	\ed 
The former is familiar in the Rubin causal model: a potential outcome---an outcome under a potential, counterfactual condition, being $0$ or $1$. The latter is a real number in the unit interval $[0, 1]$: the probability of obtaining a certain outcome under a potential condition. So it is called a {\em stochastic potential outcome} in statistics, although it is more commonly known as a {\em counterfactual probability} in philosophy. This stochastic potential outcome is essentially the parameter of a Bernoulli distribution, so let's denote it by a Greek letter, representing a parameter whose value is unknown, following the usual convention in statistics. More generally, we have:
	\opp 
	{\bf Step 2 (Stochasticize Potential Outcomes).} Replace the potential outcomes in the above by stochastic potential outcomes:
\begin{eqnarray*}
	\Cure_i^{\take=1} & \to & \kappa_i^{\take=1}
\\
	\Cure_i^{\take=0} & \to & \kappa_i^{\take=0}
\\
	\Take_i^{\assign=1} & \to & \tau_i^{\assign=1}
\\
	\Take_i^{\assign=0} & \to & \tau_i^{\assign=0} \,.
\end{eqnarray*}
Concepts originally defined in terms of the variables on the left will need to be redefined in terms of those on the right.
	\edd  
The Greek letter $\kappa$ stands for \emph{c}---i.e., the probability of getting \emph{c}ured. Similarly, the Greek letter $\tau$ represents \emph{t}---i.e., the probability of \emph{t}aking the treatment. So, for example, $\tau_i^{\assign=1}$ denotes the probability that there would be for $i$ to take the treatment if $i$ were assigned to the treatment group. For simplicity, it is assumed that each variable is binary, with only two possible outcomes. If, instead, there are $n$ outcomes with $n > 2$, the parameters on the right need to be vector-valued, existing in an $(n-1)$-dimensional space (or simplex), to describe the probabilities of all the $n$ possible outcomes under a potential condition. 

Step 2 is not new. The replacement of potential outcomes with stochastic potential outcomes was previously proposed by Greenland (1987) and Robins \& Greenland (1989) in another context, specifically to address issues in hazard analysis. They sometimes use the term `probabilities of causation' instead of `stochastic potential outcomes'. Step 2 extends this idea to the current context of instrumental variable estimation, following the suggestion of Robins \& Greenland (2000).

Now, each individual $i$ has an individual treatment effect, which is redefined using the new notation as follows:
	\begin{eqnarray}
	{\rm ITE}_i & =_\textrm{df} & \kappa_i^{\take=1}- \kappa_i^{\take=0} \,.
	\label{def-first}
	\end{eqnarray}
This represents the change in the probability of $i$'s getting cured when the counterfactual condition (indicated by the superscript) switches from not taking to taking the treatment. In the special case of the deterministic setting, those $\kappa$-probabilities are restricted to be $0$ or $1$, which limits ${\rm ITE}_i$ to values of $-1$, $0$, or $1$. But in general, ${\rm ITE}_i$ is allowed to range from $-1$ to $1$. Similarly, each individual $i$ has a degree of compliance defined by:
	\begin{eqnarray}
	{\rm DC}_i & =_\textrm{df} & \tau_i^{\assign=1} - \tau_i^{\assign=0} \,.
	\end{eqnarray}
This is the change in the probability of $i$'s taking the treatment when the counterfactual condition switches from the control to the treatment group. Compliers are now defined as those who have a positive degree of compliance; defiers, negative:
	\begin{eqnarray}
	{\rm DC}_i \;  
	\begin{cases}
		\, > 0: & \mbox{$i$ is a complier;}
	\\
		\, = 0: & \mbox{$i$ is an indifferent taker;}
	\\
		\, < 0: & \mbox{$i$ is a defier.} 
	\end{cases}
	\end{eqnarray}
The compliers defined in the classic, deterministic setting are only a special case of compliers: those with a maximal degree of compliance, ${\rm DC}_i = 1$. Consider the subpopulation consisting of all compliers:
	\begin{eqnarray}
	\text{Com} &=_\text{df}& \{i: \text{DC}_i > 0\} \,.
	\end{eqnarray}
For this subpopulation, let's define a new local average treatment effect as a weighted average, where the weights are proportional to degrees of compliance $\text{DC}_i$:
	\begin{eqnarray}
	\text{DATE}
	&=_\text{df}& \displaystyle
	\sum_{i \in \text{Com}} 
	\underbrace{
			\left( 
			\frac{{\rm DC}_i}{\sum_{j \in \text{Com}} {\rm DC}_j} 
			\right)
	}_{
		\text{weight  of $i$} \vphantom{\frac{1}{1}}
	}
	{\rm ITE}_i \,. \label{def-last}
	\end{eqnarray}
The denominator $\sum_{j \in \text{Com}} {\rm DC}_j$ is introduced only to ensure that the weights sum to $1$, thereby defining a weighted average. The term `DATE' is short for the Degree-of-compliance-weighted Average Treatment Effect. In the deterministic case, where stochastic potential outcomes (the $\kappa$- and $\tau$-probabilities) are restricted to $0$ or $1$, the DATE degenerates to the LATE. 

This definition of the DATE is similar to the weighted average treatment effect defined in Small et al.\ (2017). However, Small et al.\ (2017) still assume a joint distribution over the actuals and the simple counterfactuals in a stochastic setting---thus assuming the unique-parallel-universe view. It is this metaphysical assumption that I propose to remove, and it means that I have to work without such a joint probability distribution.\footnote{
	To be more precise, Small et al.\ (2017) define a Strength-of-IV Weighted Average Treatment Effect, or SIV-WATE, which is similar to the DATE defined here. But their definition has two properties I wish to avoid here. First, their definition is for a weighted average of {\em settled} individual treatment effects, still under the assumption of the unique-parallel-universe view. Indeed, they still stick to potential outcomes $Y(1)$ and $Y(0)$, which correspond to $\Cure^{\take=1}$ and $\Cure^{\take=0}$ in the present notation, rather than using stochastic potential outcomes. Second, although the weights in their definition are meant to capture degrees of compliance, they are defined in terms of the joint probability distribution over observables. But I believe that conceptual clarity is gained if degrees of compliance are defined directly in counterfactual terms---by stochastic potential outcomes, as proposed in this paper.
} 

The key lies in the next step. Once stochastic potential outcomes are in place, they must be connected to probabilities over observables to make observational studies possible. This requires explicitly stating an assumption that serves as a {\em bridge principle}, connecting the potential to the observable. I propose that this bridge principle is most conveniently formulated using a causal Bayes net. Indeed, a defining feature of a causal Bayes net---its factorization rule---functions as such a bridge principle, linking stochastic potential outcomes to probabilities of observables. Let me explain. 

\begin{figure}[ht]
	\centering \includegraphics[width=0.9\textwidth]{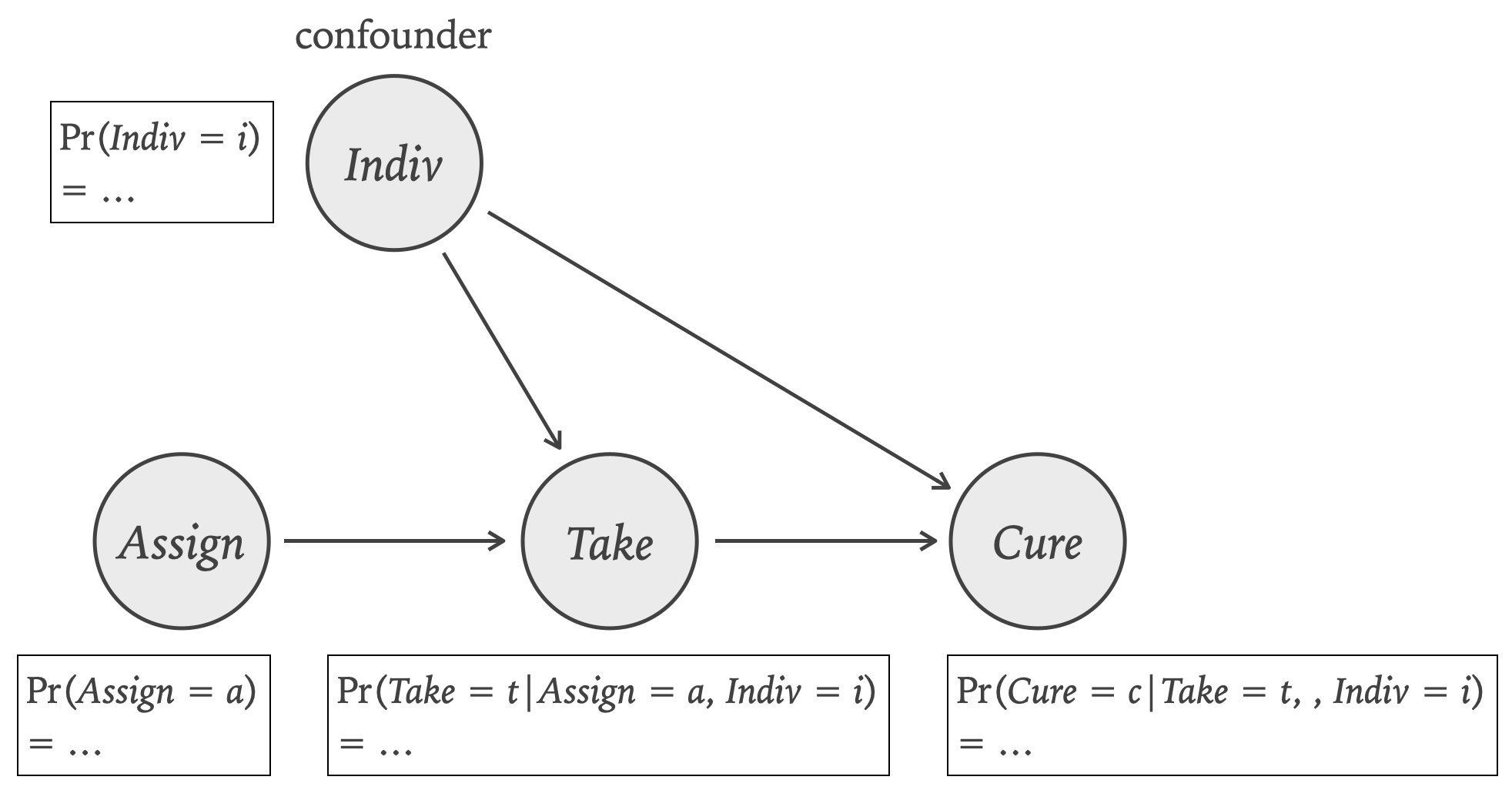}
	\caption{\em A causal Bayes net for instrumental variable estimation}
	\label{fig-cbn}
	\end{figure}

Consider the DAG (directed acyclic graph) in figure \ref{fig-cbn}, which is a very natural choice for representing the causal structure among the variables of interest.
The confounding variable, $\Individual$, represents the individual randomly chosen from the population under study. Accordingly, $\Cure$ denotes whether the randomly chosen person is cured or not. The design of this confounding variable is intended to make it very fine-grained, capturing all confounding factors (at least at a slice of time, as the present setup still does not model change over time): when $\Individual = i$ (indicating the $i$-th individual), it is understood to encompass all relevant background conditions for $i$---financial status, personal preference, health condition, and so on---regardless of whether we can identify those conditions.

To turn such a causal structure into a causal Bayes net, we only need to specify a joint probability distribution that adheres to the {\em factorization rule}:
$$
\begin{array}{lllll}
	\textrm{Joint distribution} \;\, \PP 
	\big(
		\underbrace{V_1, V_2, \ldots, V_n}_\textrm{all variables}
	\big)
\\
	= \; \displaystyle \prod_{k=1}^{n} \,\PP \big( V_k \given \textrm{the variables being the direct causes of } V_k  \big).
\end{array}
$$
Under the causal interpretation of the Bayes net, the right side actually involves counterfactual probabilities: the conditional probability of $V_k = v_k$ given its direct causes, as a parameter of the causal Bayes net, is set to be a counterfactual probability: the probability $p$ such that, if the direct causes of $V_k$ took such and such values, there would be a chance $p$ of $V_k = v_k$. I propose that such counterfactual probabilities (or parameters) in a causal Bayes net are just stochastic potential outcomes.  

For instance, using the DAG in figure \ref{fig-cbn}, the conditional probability of $\Cure = 1$ given that its direct causes take such and such values is equal to the following counterfactual probability:
\begin{eqnarray*}
	\PP \big( \Cure = 1 \mid  
	\underbrace{
		\Take = 1, \Individual = i
	}_{
		\text{direct causes of $\Cure$}
	}
	\big)
		&=& \kappa_i^{\take=1} \,.
\end{eqnarray*}
So, under the causal interpretation of the Bayes net, the full meaning of a factorization rule can be schematized like this:
$$
\begin{array}{lll}
	\PP\big( 
	\overbrace{
		\Cure = 1, \Take = 1, \Assign = 1,  
	}^{
		\textbf{observable}
	}
	\Individual = i
	\big)
\\[0.5em]
	=\; \PP \left( \Cure = 1 \given \Take = 1, \Individual = i \right)
	& \leftarrow & \quad\;\, \kappa_i^{\take=1} 
\\
	\quad\; \cdot \; \PP \left( \Take = 1 \given \Assign = 1, \Individual = i \right)
	& \leftarrow & 
	\underbrace{
		\tau_i^{\assign=1}
	}_{
	\quad\textbf{counterfactual}
	}
\\[-1.2em]
	\quad\; \cdot \; \PP \left( \Assign = 1 \right)
\\
	\quad\; \cdot \; \PP \left( \Individual = i \right) \,,
\end{array}
$$
where the `$\leftarrow$' notation means plugging the value on the right into the left (as in the programming language \emph{R}). For another instance of the factorization rule:
$$
\begin{array}{lll}
	\PP\big( 
	\overbrace{
		\Cure = 1, \Take = 0, \Assign = 0,  
	}^{
		\textbf{observable}
	}
	\Individual = i
	\big)
\\[0.5em]
	=\; \PP \left( \Cure = 1 \given \Take = 0, \Individual = i \right)
	& \leftarrow & \kappa_i^{\take=0} 
\\
	\quad\; \cdot \; \PP \left( \Take = 0 \given \Assign = 0, \Individual = i \right)
	& \leftarrow & 
	\underbrace{
		1-\tau_i^{\assign=0}
	}_{
	\textbf{counterfactual}
	}
\\[-1.2em]
	\quad\; \cdot \; \PP \left( \Assign = 1 \right)
\\
	\quad\; \cdot \; \PP \left( \Individual = i \right) \,.
\end{array}
$$
These are bridge principles that connect the counterfactual to the observable.

Or, more formally, the factorization rule, when applied to the present DAG, states the following:
\begin{eqnarray}
	&& \PP\left( \Cure = c, \Take = t, \Assign = a, \Individual = i \right) 
	\nonumber
\\[0.2em]
	&& =\; \PP \left( \Cure = c \given \Take = t, \Individual = i \right)
	\label{for-factorization}
\\[-0.2em]
	&& \quad\; \cdot \; \PP \left( \Take = t \given \Assign = a, \Individual = i \right)
	\nonumber
\\[-0.2em]
	&& \quad\; \cdot \; \PP \left( \Assign = a \right)
	\nonumber
\\[-0.2em]
	&& \quad\; \cdot \; \PP \left( \Individual = i \right)
	\nonumber
\end{eqnarray}
The conditional probabilities on the right---{\em qua} probabilities over one variable or another given its direct causes---are essentially counterfactual probabilities, as required by the causal interpretation of the Bayes net. This can be made explicit by the following bridge principle:
\begin{eqnarray}
	\PP \left( \Cure = 1 \mid  \Take = 1, \Individual = i\right) 
		&=& \kappa_i^{\take=1}
	\nonumber
\\
	\PP \left( \Cure = 1 \mid  \Take = 0, \Individual = i\right)		
		&=& \kappa_i^{\take=0}
	\nonumber
\\
	\PP \left( \Take = 1 \mid  \Assign = 1, \Individual = i\right)
		&=& \tau_i^{\assign=1}
	\label{for-plug-in}
\\
	\PP \left( \Take = 1 \mid  \Assign = 0, \Individual = i\right) 
		&=& \tau_i^{\assign=0}
	\nonumber
\\
	\PP \left( \Individual = i\right) 
		&=& \tfrac{1}{N}	 
	\nonumber
\\
	\PP \left( \Assign = 1\right) 
		&=& \tfrac{1}{2} \,,
	\nonumber
\end{eqnarray}
where $N$ denotes the size of the entire population. The first four lines incorporate counterfactual probabilities as parameters of the causal Bayes net. The last two lines describe a randomization mechanism: all individuals in the population have equal probabilities of being chosen, and equal probabilities of being assigned to the treatment or control group. This illustrates the idea of the third and final step in the stochastic update to the potential outcome framework:
	\opp 
	{\bf Step 3 (Incorporate Stochastic Potential Outcomes into a Causal Bayes Net).} Plug the stochastic potential outcomes from the previous step into (the right side of the factorization rule of) an appropriate causal Bayes net. This provides a bridge principle connecting the counterfactual to the observable, thereby making observational studies possible.  
	\edd 
So, to recap, the potential outcome framework can be made fully stochastic in three steps: the potential outcomes in use are first identified (in Step 1) and then replaced by stochastic potential outcomes (in Step 2), which are then plugged into an appropriate causal Bayes net (Step 3).  

It is time to state the theorem that meets the challenge. I proved it elsewhere (Lin 2025 manuscript), in a paper written for philosophers and discussing issues about not metaphysics, but the interaction between (i) the logic of deductive inference and (ii) the theory of causal inference as a kind of non-deductive inference. For self-containment, a proof is provided in the Appendix.

\begin{theorem}[Lin 2025 manuscript]
	Under the following four assumptions:
	\op 
	\im[$\mathrm{(i)}$] no defiers $(\mathrm{DC}_i < 0 \text{ for no }i)$, 
	\\[-2em]
	\im[$\mathrm{(ii)}$] existence of compliers $(\mathrm{DC}_i > 0 \text{ for some }i)$, 
	\\[-2em]
	\im[$\mathrm{(iii)}$] the factorization rule $(\ref{for-factorization})$ for the causal Bayes net in figure $\ref{fig-cbn}$, 
	\\[-2em]
	\im[$\mathrm{(iv)}$] the bridge principle $(\ref{for-plug-in})$, 
	\ed
	the {\rm DATE} is equal to the usual {\rm IV} estimand:
	\begin{eqnarray*}
	{\rm DATE}
	&=& \frac{
	\PP\left(\Cure = 1\given \Assign = 1\right) \,-\, \PP\left(\Cure = 1 \given \Assign = 0\right)
	}{
	\PP\left(\Take = 1 \given \Assign = 1\right) \,-\, \PP\left(\Take = 1 \given \Assign = 0\right)
	} \,.
	\end{eqnarray*}
\end{theorem}
In this theorem, treatment outcomes are restricted to be binary. Extension of this result to multi-valued or continuous valued treatment outcomes is, I believe, straightforward and will be included in an updated version of this paper. For now, the focus is on how this theorem addresses some metaphysical and methodological issues about causal inference.

Note that this theorem does not assume the unique-parallel-universe view. In particular, there is no single joint probability distribution over the factual variables, the stochastic counterfactuals, and the simple counterfactuals. There is only a joint probability distribution over the factual variables, as expressed by the factorization rule, together with a Bernoulli distribution for each stochastic potential outcome in use ($\tau_i^{\cdots}$ or $\kappa_i^{\cdots}$). Those distributions are not assumed to be marginal distributions of a single, common, big joint probability distribution. That is, those distributions live in separate probability spaces, with their own sample spaces, containing only coarse-grained possibilities. So, metaphysically, there is no use of fine-grained possibilities, and thus no commitment to the unique-parallel-universe view. 

Mathematically, the identification result can be proved without assuming a single, common, big joint probability distribution. To be sure, one is free, if one wishes, to construct such a single, large-scale joint probability distribution to house the separate, small-scale probability distributions assumed in the theorem, and to treat it instrumentally as a mere mathematical construction without any underlying metaphysical view. But this mathematical construction will have to come with decisions about how the variables involved are dependent or independent of each other; and such a mathematical construction and its associated decisions are entirely irrelevant to the truth of the theorem stated above. The issue of the dependence or independence among the factual variables and the simple counterfactuals, which has often been discussed in the classic, deterministic literature, is entirely irrelevant to the present result. 

In the special case of the deterministic setting, all compliers have a maximal degree of compliance $\text{DC}_i \equiv 1$. Consequently, the weights in the weighted average become equal, the DATE reduces to the familiar LATE, and the above theorem reproduces the classic identification result for the LATE: 
	\begin{eqnarray*}
	{\rm LATE}
	&=& \frac{
	\PP\left(\Cure = 1\given \Assign = 1\right) \,-\, \PP\left(\Cure = 1 \given \Assign = 0\right)
	}{
	\PP\left(\Take = 1 \given \Assign = 1\right) \,-\, \PP\left(\Take = 1 \given \Assign = 0\right)
	} \,.
	\end{eqnarray*}
The existing practice of instrumental variable estimation can thus be reinterpreted: it was previously understood to estimate the LATE in the deterministic setting, but has actually been estimating a more general quantity all along---the DATE---in a general stochastic setting, with the deterministic setting as merely a special, degenerate case, and without assuming excessive metaphysics: in particular, without assuming the unique-parallel-universe view.

%\end{document}

\section{Varieties of Causal Models}\label{sec_challenges}

I take the account developed above to be a synthesis of two types of causal models: the Rubin causal model and causal Bayes nets. I will elaborate on this claim below, and defend the use of graphical causal models in general and causal Bayes nets in particular---against two familiar challenges.

\subsection{Challenge from Rubin Causal Modelers (Imbens)}\label{sec_Imbens}

Imbens (2020) argues that the Rubin causal model is superior to graphical (DAG-based) causal models for an important task: facilitating the statement of various assumptions underlying causal inference. I look at the matter differently: the Rubin causal model and the causal Bayes net excel at stating different types of assumptions, so they deserve a synthesis. 

To see why, let's start with what Imbens (2020) considers a prime example, the assumption of no defiers (also called monotonicity):
	\opp 
	({\em No Defiers, Deterministic Version})
	
	$
	\Take_i^{\assign=1} \;\ge\; \Take_i^{\assign=0}
	$
	for all $i$.
	\edd 
In this case, DAGs are not needed, nor do they make it easier to state the assumption. Even in the stochastic update developed above, DAGs are still not necessary to state the no-defiers assumption: $\text{DC}_i =_\text{df} \tau^{\assign=1} - \tau^{\assign=0} < 0$ for no $i$. Or equivalently:
	\opp 
	({\em No Defiers, Stochastic Version})
	
	$
	\tau_i^{\assign=1} \;\ge\; \tau_i^{\assign=0}
	$ 
	for all $i$.
	\edd 
So Imbens' point is well-taken for this example. 

The use of DAGs is also not particularly helpful for stating another type of assumption: the assumption of no interference between individuals. This assumption greatly simplifies causal inference---it allows us to ignore the counterfactual conditions (superscripts) that refer to multiple people simultaneously. Under the no-interference assumption, we no longer need to consider multi-person potential outcomes such as $\Cure_i^{\take_i=1, \take_j =0}$, which denotes the medical outcome that $i$ would have if $i$ took the treatment but {\em another} individual $j$ did not. Instead, we can safely proceed with single-person potential outcomes such as $\Cure_i^{\take_i =1}$, and even simplify the notation as $\Cure_i^{\take =1}$ without confusion. To be sure, it is possible to draw a DAG to express the assumption of no interference, but that would be quite inconvenient---it would require a very big DAG, which includes many smaller separated DAGs, as many as there are individuals in the population. 

However, there is a type of assumption that seems to me most naturally stated by incorporating the style of a causal Bayes net into the mix. To reiterate a point I made in the previous section: to make observational studies possible, a bridge principle needs to be assumed to connect the counterfactual to the observable. This is no exception in the deterministic setting formalized by the Rubin causal model, in which we often see bridge principles like the following:
	\opp 
	({\em Consistency}) The probability distribution over a potential outcome $Y^{x=...}$ is equal to that over the corresponding observable $Y$ given the truth of the counterfactual condition. For example: 
	\begin{eqnarray*}  
		\PP \big( \, 
		\underset{
			\textbf{counterfactual} \quad\quad
		}{
			\underbrace{Y^{x=1}  = 1} \given X=1
		}
		\big)
		&=& 
		\PP \big( \!
		\underset{
			\textbf{observable} \quad\quad\quad
		}{
			\underbrace{Y  = 1} \given X=1 
		}
		\!\! \big)
		\,.
	\end{eqnarray*}
	({\em Randomization}) The observable variable $\Assign$ is probabilistically independent of the set of all the four potential outcomes in use,
	$\Take^{\assign = 0}$,
	$\Take^{\assign = 1}$,
	$\Cure^{\take = 0}$, and
	$\Cure^{\take = 1}$; or in symbols:
	\begin{eqnarray*}
	&&
		\PP
		\Big(
		\overbrace{
		\Take^{\assign = 0} = t, \,
		\Take^{\assign = 1} = t', \,
		\Cure^{\take = 0} = c, \,
		\Cure^{\take = 1} = c' 
		}^{
			\textbf{counterfactual}
		}
		\Big)
	\\
	&&=\;
		\PP
		\Big(
		\Take^{\assign = 0} = t, \,
		\Take^{\assign = 1} = t', \,
		\Cure^{\take = 0} = c, \,
		\Cure^{\take = 1} = c' \,\Big|\, 
		\underbrace{
			\Assign = a
		}_{
			\textbf{observable}
		} 
		\Big)
	\end{eqnarray*}
	\edd   
Both of those assumptions are made in the classic result about the LATE. When we move onto the stochastic setting and
replace potential outcomes by stochastic potential outcomes (with $\Take$ replaced by $\tau$ and $\Cure$ by $\kappa$), the above two assumptions need to be rewritten. A brute-force rewriting would introduce some complexity: we get probabilities of probabilities, or to be more precise, a joint probability distribution over stochastic potential outcomes. Fortunately, we can avoid such complexity by appeal to a causal Bayes net. Indeed, we have seen that the causal Bayes net in the above allows us to state a new bridge principle---a factorization rule with stochastic potential outcomes as plug-ins, schematized as follows: 
$$
\begin{array}{lll}
	\PP\big( 
	\overbrace{
		\Cure = 1, \Take = 1, \Assign = 1,  
	}^{
		\textbf{observable}
	}
	\Individual = i
	\big)
\\[0.5em]
	=\; \PP \left( \Cure = 1 \given \Take = 1, \Individual = i \right)
	& \leftarrow & \quad\;\, \kappa_i^{\take=1} 
\\
	\quad\; \cdot \; \PP \left( \Take = 1 \given \Assign = 1, \Individual = i \right)
	& \leftarrow & 
	\underbrace{
		\tau_i^{\assign=1}
	}_{
	\quad\textbf{counterfactual}
	}
\\[-1.2em]
	\quad\; \cdot \; \PP \left( \Assign = 1 \right)
\\
	\quad\; \cdot \; \PP \left( \Individual = i \right) \,.
\end{array}
$$
The use of such bridge principles---along with the causal Bayes net they presuppose---avoids the complexity inherent in the probabilities of probabilities. Moreover, this setting is quite flexible: it is mainly designed for stochastic cases but still allows for the special case where all stochastic potential outcomes are $0$ or $1$, which brings back determinism at the individual level. 

So, while the Rubin causal model is more convenient for stating the assumptions exclusively about potential outcomes and even stochastic potential outcomes, causal Bayes nets are more convenient for stating bridge principles that connect the counterfactual to the observable at least in the stochastic setting. 

The proposed account---a stochastic update to the Rubin causal model---is truly a synthesis of the Rubin causal model with a causal Bayes net, allowing us to enjoy the best of both worlds.

\subsection{Challenge from Structural Equation Modelers (Pearl)}\label{sec_Pearl}

Although Pearl (2009) is one of the main proponents of graphical causal models, he opts for {\em structural equation models}, which, from a mathematical point of view, are a special case of causal Bayes nets, restricting the probabilities over endogenous variables to be $0$ or $1$. That amounts to imposing determinism within a causal Bayes net. Pearl sought to justify his preference for structural equation models with two claims (Pearl 2009: section 1.4.4):
	\op 
	\xm {\em Pearl's Claim $A$}. The deterministic assumption provides enough resources to graphical causal models for defining an important concept: the concept of counterfactual probabilities or probabilities of counterfactuals given the available evidence. Causal Bayes nets lack the resources needed to provide such a definition.
	
	\xm {\em Pearl's Claim $B$}. Although the definition sought in Claim $A$ is achieved under the deterministic assumption, there is no loss of generality. The joint distribution of an (indeterministic) causal Bayes net over a variable set $\VV$ can always be recovered by constructing a structural equation model over a larger variable set $\VV \cup \UU$, which adds to each variable in $\VV$ an exogenous parent (direct cause).
	\ed 
I believe that the relative advantage promised in Claim $A$ is actually mistaken. If so, the addition of variables in Claim $B$ only increases the mathematical complexity of causal modeling and brings back the metaphysical baggage along with the deterministic assumption, without gaining the relative advantage promised in Claim $A$. Let me explain. 

Pearl believes that, in causal modeling, it is of utmost importance to define the concept of probabilities of counterfactuals given the available evidence, which are answers to queries like the following: 
	\opp 
	({\em Query about Probabilities of Counterfactuals})
	
	What is the probability that $[$ individual $i$ would have been cured had $i$ taken the treatment $]$, given the evidence that $i$ actually failed to be cured and developed such and such symptoms?
	\edd 
Pearl shows how such a probability can be defined in a given structural equation model, and provides reasons for thinking that this cannot be done with an (indeterministic) causal Bayes net. But Pearl's starting point---the form of the queries of interest---already presupposes the unique-parallel-universe view. In the language of the Rubin causal model, Pearl's query asks for probabilities of potential outcomes:
	\opp 
	({\em Query about Probabilities of Potential Outcomes}) 
	
	What is the probability of $\Cure_i^{\take=1} = 1$, given the evidence that $i$ actually failed to be cured and developed such and such symptoms?
	\edd 
But this presupposes a joint probability distribution over the actual variables and counterfactual variables, and thus makes a strong metaphysical assumption: the unique-parallel-universe view. 

Here is my proposal for addressing that problem. We have seen that, in the stochastic setting, the talk of potential outcomes needs to be replaced by talk of stochastic potential outcomes. So, a better way to frame the query is this:
	\opp 
	({\em Query about Potential Probabilities}) What would have been the probability $p$ for individual $i$ to be cured had $i$ taken the treatment, given the evidence that $i$ actually failed to be cured and developed such and such symptoms? 
	\edd 
This asks about our estimate of $p$ as a counterfactual probability, or the value of a stochastic potential outcome, given the available evidence. More generally, we can ask questions of this form:
	\opp 
	What would be the probability for something to happen were the value of $X$ set to be $x$, given evidence $\mathcal{E}$? Let the answer be encoded by a probability distribution written $\PP\left(\,\cdot \GGiven \mathtt{do}(X=x) \Given \mathcal{E}\right)$.
	\edd
Let me show how this can be defined in a given causal Bayes net. 

Start with a special case of the above query, in which the evidence is complete, assigning values to all variables in a given causal Bayes net (all variables in $\VV$):
	\opp 
	What would be the probability for something to happen were the value of $X$ set to be $x$, given the {\em complete} evidence $\VV = \vv$? Let the answer be encoded by a probability distribution written $\mathbb{P}_{\,\vv}^{\,x}\left( \,\cdot\, \right)$.
	\edd
The subscript and superscript follow this notational convention: the subscript denotes the actual, the superscript denotes the potential or counterfactual. This stochastic potential outcome can be defined in a given causal Bayes net that comes with a joint distribution $\PP(\,\cdot\,)$ over the actual variables and a DAG:
\begin{eqnarray*}
	\mathbb{P}_{\,\vv}^{\,x}\left( \,\cdot\, \right)
&=_\text{df}& 
	\PP\left(\,\cdot \Given X=x, (\VV=\vv) \!\!\upharpoonright_{\text{NonDes}[X]} \right) \,,
\end{eqnarray*}
where $(\VV=\vv) \!\!\upharpoonright_{\text{NonDes}[X]}$ denotes the restriction of the value assignment $\VV = \vv$ to the non-descendants of $X$. Allowing evidence $\mathcal{E}$ to be incomplete (represented by a set of possible value assignments to the variables in $\VV$), our estimate of the stochastic potential outcome given $\mathcal{E}$ can be defined as a weighted average of stochastic potential outcomes given different bodies of complete evidence: 
\begin{eqnarray*}
	\PP\left(\,\cdot \GGiven \mathtt{do}(X=x) \Given \mathcal{E}\right)
&=_\text{df}& 
	\sum_{\vv} \,
	\PP_{\,\vv}^{\,x}\left( \,\cdot\, \right) \, \PP\left( \VV = \vv \given \mathcal{E} \right) \,.
\end{eqnarray*}
This is a definition set within the framework of causal Bayes nets, without the unique-parallel-universe assumption. 

Lesson: once we clearly distinguish between potential outcomes, stochastic potential outcomes, and the different queries they prompt, it will not be hard to unleash more expressive power from causal Bayes nets. 

So, the relative advantage promised in Pearl's Claim $A$ is mistaken. There is thus no relative advantage to justify the mathematical and metaphysical costs incurred by Pearl's addition of variables, which undermines Pearl's Claim $B$. When we wish to use graphical causal models in a fully stochastic setting, causal Bayes nets remain a good choice---better than the alternatives considered above.

\section{Conclusion}\label{sec_conclusion}

I have proposed a stochastic update to the potential outcome framework and generalized the classic LATE identification result without taking on the metaphysical commitments tacit in its standard stochastic interpretation. The diagnosis was that the standard joint-probability definitions of compliers and the LATE, once interpreted stochastically, presuppose a substantive Laplacian thesis---the unique-parallel-universe assumption---that Dawid (2000) finds implausible and Lewis (1973) argues to be incoherent. The remedy was to accept only the weaker commonsense metaphysics, to assign probabilities only to coarse-grained possibilities, and to bridge counterfactual quantities to observables via the factorization rule of a causal Bayes net. Within this framework, the Degree-of-compliance-weighted Average Treatment Effect (DATE) generalizes the LATE, and the reported theorem shows that under assumptions analogous to those used in the classic result, the DATE equals the usual IV estimand. In the deterministic special case, the DATE reduces to the LATE.

Three implications are worth flagging. First, existing IV practice can be reinterpreted: what has long been understood as estimating the LATE in the deterministic setting has in fact been estimating the DATE in a more general setting: the stochastic setting, with the deterministic case as a degenerate special case. Second, the proposed framework is a genuine synthesis. It retains the Rubin causal model's transparent statement of substantive assumptions about individual-level potential outcomes and stochastic potential outcomes---such as no defiers and no interference---while exploiting the causal Bayes net's transparent statement of a bridge principle from the counterfactual to the observable. Third, the synthesis does not require the unique-parallel-universe assumption inherent in Pearl's structural equation models.

The present work may amount to more than just an extension of the classic LATE result. The three steps that yielded the DATE in Section~\ref{sec_late} seem to suggest a general recipe for moving the classic potential outcome framework to a fully stochastic setting:
\op
\im {\bf Step 1 (Identify).} Identify the (individual-level) potential outcomes appearing in the deterministic statement of the result of interest---in our case, $\Cure_i^{\take=1}$, $\Cure_i^{\take=0}$, $\Take_i^{\assign=1}$, and $\Take_i^{\assign=0}$.
\im {\bf Step 2 (Stochasticize).} Replace each potential outcome by its stochastic counterpart, such as a Bernoulli parameter recording the chance a binary outcome would have under the corresponding counterfactual condition---and redefine the relevant concepts (individual treatment effects, compliance types, target estimands) in these stochastic terms. In the LATE case, this is how the LATE gave way to the DATE.
\im {\bf Step 3 (Bridge via a Causal Bayes Net).} Embed the stochastic potential outcomes as parameters of an appropriate causal Bayes net by plugging them into the right-hand side of the factorization rule. This supplies the bridge principle that connects the counterfactual to the observable and makes observational studies possible---makes an identification result possible in a fully stochastic setting.
\ed
The classic deterministic result is then recovered as a degenerate special case by restricting all stochastic potential outcomes to the values $0$ or $1$.

Several questions remain open. The four assumptions in the main theorem play distinct roles, and a more careful analysis of how they relate to the assumptions of the classical framework would clarify the conceptual structure of the proposed account. In particular, it would be useful to determine which of the four plays the role of a stochastic analogue of the classic Stable Unit Treatment Value Assumption (SUTVA), and which plays the role of an analogue of the classic consistency and randomization assumptions. A natural further question is whether the three-step recipe above generalizes other classic identification results in the Rubin tradition---by combining stochastic potential outcomes with an appropriately chosen causal Bayes net. I leave these for future work.

%\end{document}

\appendix

\section*{Appendix: Proof of the Main Result (Theorem 1)}\label{appendix}

The goal is to verify the following equation:
	\begin{eqnarray*}
	{\rm DATE}
	&\overset{?}{=}&
	\frac{
		\PP\left( \Cure = 1 \given \Assign = 1  \right)
		\,-\, \PP\left( \Cure = 1 \given \Assign = 0  \right)
	}{
		\PP\left( \Take = 1 \given \Assign = 1  \right)
		\,-\, \PP\left( \Take = 1 \given \Assign = 0  \right)
	} \,.
	\end{eqnarray*}
The terms on the right-hand side are to be calculated in turn. I will leverage an equivalent formulation of the factorization rule, called the {\em causal Markov condition}: every variable in the DAG is probabilistically independent of its non-descendants (non-effects) given its parents (direct causes). For readability, the four parameters $\kappa_i^{\take=1}$, $\kappa_i^{\take=0}$, $\tau_i^{\assign=1}$, and $\tau_i^{\assign=0}$ will be abbreviated as $\kappa_i^{1}$, $\kappa_i^{0}$, $\tau_i^{1}$, and $\tau_i^{0}$, respectively.

Start with the first term in the numerator:
	\begin{align*}
	&\PP\left( \Cure = 1 \given \Assign = 1  \right)
\\[0.8em]
	&=\;
		\sum_{i,t} \Big(
		\PP \,( 
		\Cure = 1 \given 
		\Take = t, \Individual = i, 
		\Assign = 1
		) 
	\\[-0.8em]
	& \quad\quad\quad\quad\cdot\;  
		\PP\left( 
		\Take = t \given 
		\Individual = i, \Assign = 1
		\right)
	\\
	& \quad\quad\quad\quad\cdot\;  
		\PP\,(  
		\Individual = i \given 
		\Assign = 1
		)
	\Big)
	& \mbox{by Chain Rule}
\\[0.4em]
	&=\;
		\sum_{i,t} \Big(
		\PP \,( 
		\Cure = 1 \given 
		\Take = t, \Individual = i, 
			\cancel{\Assign = 1}
		) 
	\\[-0.8em]
	& \quad\quad\quad\quad\cdot\;  
		\PP\left( 
		\Take = t \given 
		\Individual = i, \Assign = 1
		\right)
	\\
	& \quad\quad\quad\quad\cdot\;  
		\PP\,(  
		\Individual = i \given 
			\cancel{\Assign = 1}
		)
	\Big)
	& \mbox{by Causal Markov}
\\[0.4em]
	&=\;
		\sum_{i} \Big(
		\PP\left( 
		\Cure = 1 \given 
		\Take = 1, \Individual = i  
		\right) 
	\\[-0.8em]
	& \quad\quad\quad\quad\cdot\;  
		\PP\left( 
		\Take = 1 \given 
		\Individual = i, \Assign = 1
		\right)
	\\
	& \quad\quad\quad\quad\cdot\;  
		\PP\left(  
		\Individual = i 
		\right)
	\Big)
	\\[-0.2em]
	&
		\quad\;\;+\;\sum_{i} \Big(
		\PP\left( 
		\Cure = 1 \given 
		\Take = 0, \Individual = i  
		\right) 
	\\[-0.8em]
	& \quad\quad\quad\quad\cdot\;  
		\PP\left( 
		\Take = 0 \given 
		\Individual = i, \Assign = 1
		\right)
	\\
	& \quad\quad\quad\quad\cdot\;  
		\PP\left(  
		\Individual = i
		\right)
	\Big)
	%&& \mbox{just rearrangement}
\\[0.8em]
	&=\; \sum_{i} \Big( \kappa_i^{1}\, \tau_i^{1} \, \tfrac{1}{N} \Big) \;+\;  \sum_{i} \Big(\kappa_i^{0} \, (1 - \tau_i^{1}) \, \tfrac{1}{N} \Big) 
	& \mbox{by Bridge Principle (\ref{for-plug-in})}
	%&& \mbox{plugging in parameters}
	\\[0.5em]
	&=\; \tfrac{1}{N} \sum_{i} \Big( \kappa_i^{1}\, \tau_i^{1}  + \kappa_i^{0} \, (1 - \tau_i^{1}) \Big) \,.
	%&& \mbox{just a bit algebra}
	\end{align*}
%% Note how the Causal Markov Condition is applied here. It suggests that, the proof is valid as long as (i) the parents of $\Cure$ are exactly $\Take$ and $\Individual$ and (ii) the assignment variable $\Assign$ is a non-descendant of $\Cure$. 
Similarly for the second term in the numerator:
	\begin{align*}
	&\PP\left( \Cure = 1 \given \Assign = 0  \right)
	\\
	&=\; \tfrac{1}{N} \sum_{i} \Big( \kappa_i^{1}\, \tau_i^{0}  + \kappa_i^{0} \, (1 - \tau_i^{0}) \Big) \,.
	\end{align*}
Now calculate the first term in the denominator:
	\begin{align*}
	&\PP\left( \Take = 1 \given \Assign = 1  \right)
	\\
	&=\;
		\sum_{i} \Big(
		\PP\left( 
		\Take = 1 \given 
		\Individual = i, \Assign = 1  
		\right) 	
	\cdot   
		\PP\,( 
		\Individual = i 
		\given 
		\underset{
			\rm by \, Causal \, Markov
			}{
			\cancel{\Assign = 1}
			}  
		)
		\Big)
	\\
	&=\; \sum_{i} \,\tau_i^{1} \, \tfrac{1}{N}
	\\
	&=\; \tfrac{1}{N} \sum_{i} \,\tau_i^{1} \,.
	\end{align*}
%% need backtracking to generalize this step?!
Similarly for the second term in the denominator:
	\begin{align*}
	&\PP\left( \Take = 1 \given \Assign = 0  \right)
	\\
	&=\; \tfrac{1}{N} \sum_{i} \,\tau_i^{0} \,.
	\end{align*}	
To finish off, plug the four terms just calculated into the following:
	\begin{align*}
	&\frac{
		\PP\left( \Cure = 1 \given \Assign = 1  \right)
		\,-\, \PP\left( \Cure = 1 \given \Assign = 0  \right)
	}{
		\PP\left( \Take = 1 \given \Assign = 1  \right)
		\,-\, \PP\left( \Take = 1 \given \Assign = 0  \right)
	}
\\[0.5em]
	&=\;
	\frac{
		 \cancel{\tfrac{1}{N}} \sum_{i} \Big( \kappa_i^{1}\, \tau_i^{1}  + \kappa_i^{0} \, (1 - \tau_i^{1}) \Big)
		 -
		  \cancel{\tfrac{1}{N}} \sum_{i} \Big( \kappa_i^{1}\, \tau_i^{0}  + \kappa_i^{0} \, (1 - \tau_i^{0}) \Big)
	}{
		 \cancel{\tfrac{1}{N}} \sum_{i} \,\tau_i^{1}
		 -
		 \cancel{\tfrac{1}{N}} \sum_{i} \,\tau_i^{0}
	}
\\[0.5em]
	&=\;
	\frac{
		 \sum_{i} \big( \tau_i^{1} - \tau_i^{0} \big)\big( \kappa_i^{1}- \kappa_i^{0} \big)
	}{
		 \sum_{j} \big(\tau_j^{1} - \tau_j^{0}\big)
	}
\\[0.5em]
	&=\; \sum_{i} 
			\left( 
			\frac{\tau_i^{1} - \tau_i^{0}}{\sum_{j} \big(\tau_j^{1} - \tau_j^{0}\big)} 
			\right)
	\big( \kappa_i^{1}- \kappa_i^{0} \big)
\\[0.5em]
	&=\; \sum_i \left( 
			\frac{{\rm DC}_i}{\sum_{j} {\rm DC}_j} 
			\right) {\rm ITE}_i
\\[0.5em]
	&=\; \sum_{i \in \text{Com}} \left( 
			\frac{{\rm DC}_i}{\sum_{j\in \text{Com}} {\rm DC}_j} 
			\right) {\rm ITE}_i
			\quad\quad \mbox{by No Defiers}
\\[0.5em]
	&=\; {\rm DATE} \,, 
	\end{align*}
which is well-defined, since the denominator is nonzero by the existence of compliers. This completes the proof.

\section*{References}
\begin{description}
	\item Angrist, J. D., Imbens, G. W., and Rubin, D. B. (1996) ``Identification of Causal Effects Using Instrumental Variables'', {\em Journal of the American Statistical Association}, 91(434): 444-455.
	
	\item Dawid, A. P. (2000) ``Causal Inference without Counterfactuals'', {\em Journal of the American Statistical Association} 95(450): 407-424.
	
	\item Greenland, S. (1987) ``Interpretation and Choice of Effect Measures in Epidemiologic Analysis'', {\em American Journal of Epidemiology}, 125(5): 761-768.
	
	\item H\'{a}jek, A. (2014) ``Probabilities of Counterfactuals and Counterfactual Probabilities'', {\em Journal of Applied Logic}, 12(3): 235-251.
	
	\item H\'{a}jek, A. (manuscript) ``Most Counterfactuals Are False'', Research School of Social Sciences, Australian National University.
	
	\item Imbens, G. W. (2020) ``Potential Outcome and Directed Acyclic Graph Approaches to Causality: Relevance for Empirical Practice in Economics'', {\em Journal of Economic Literature} 58(4): 1129-1179.
	 
	\item Imbens, G. W. and Angrist, J. (1994) ``Identification and Estimation of Local Average Treatment Effects'', {\em Econometrica} 62, 467-476.
	
	\item Lewis, D. K. (1973) {\em Counterfactuals}, Blackwell.
	
	\item Lin, H. (2025 manuscript) ``The Logic of Counterfactuals and the Epistemology of Causal Inference'', 	arXiv:2405.11284 [cs.AI]. \url{https://arxiv.org/abs/2405.11284}
	
	\item Pearl, J. (2009), {\em Causality: Models, Reasoning, and Inference}, 2nd Edition, Cambridge University Press.
	
	%\item Pearl J., Glymour, M., and Jewell, N. P. (2016) {\em Causal Inference in Statistics: A Primer}, Chichester: Wiley.
	
	\item Robins, J. and Greenland, S. (1989) ``The Probability of Causation under a Stochastic Model for Individual Risk'', {\em Biometrics}, 45(4): 1125-1138.
	
	\item Robins, J. M. and Greenland, S. (2000) ``Comment on `Causal Inference without Counterfactuals' by A. P. Dawid'', {\em Journal of the American Statistical Association}, 95(450): 431-435.
	 
	\item Rubin, D. B. (1974) ``Estimating Causal Effects of Treatments in Randomized and Nonrandomized Studies'', {\em Journal of Educational Psychology} 66: 688-701.
	
	\item Santorio, P. (2025) ``Probabilities of Counterfactuals Are Counterfactual Probabilities'', {\em Journal of Philosophy}.
	
	\item Small, D. S., Tan, Z., Ramsahai, R. R., Lorch, S. A., and Brookhart, M. A. (2017) ``Instrumental Variable Estimation with a Stochastic Monotonicity Assumption'', {\em Statistical Science}, 32 (4), 561-579.
	
	\item Spirtes, P., Glymour, C. N., \& Scheines, R. (2000) {\em Causation, Prediction, and Search}, MIT Press.
	
	\item Stalnaker, R. C. (1968) ``A Theory of Conditionals'', in Harper, W. L., Pearce, G. A., \& Stalnaker, R. C. (eds.) {\em Ifs: Conditionals, Belief, Decision, Chance and Time}, Springer Netherlands: 41-55.
	
	\item Stalnaker, R. C. (1981) ``A Defense of Conditional Excluded Middle'', in Harper, W. L., Pearce, G. A., \& Stalnaker, R. C. (eds.) {\em Ifs: Conditionals, Belief, Decision, Chance and Time}, Springer Netherlands: 87-104.
	
	\item VanderWeele, T. J. and Robins, J. M. (2012) ``Stochastic Counterfactuals and Stochastic Sufficient Causes'', {\em Statistica Sinica}, 22(1): 379-392.
	
	\item Williams, J. R. G. (2010) ``Defending Conditional Excluded Middle'', {\em No\^{u}s}, 44(4): 650-668.
\end{description}

\end{document}